\documentclass{egpubl}
\usepackage{eg2025}
 
\ConferencePaper        %

\CGFccby

\usepackage[T1]{fontenc}
\usepackage{dfadobe}  

\biberVersion
\BibtexOrBiblatex

\electronicVersion
\PrintedOrElectronic

\ifpdf \usepackage[pdftex]{graphicx} \pdfcompresslevel=9
\else \usepackage[dvips]{graphicx} \fi

\usepackage{egweblnk} 

\title[Learning Uniformly Distributed Embedding Clusters of Stylistic Skills for Physically Simulated Characters]%
      {Learning Uniformly Distributed Embedding Clusters of Stylistic Skills for Physically Simulated Characters}

\author[]
{\parbox{\textwidth}{\centering Nian Liu$^{1,2}$, \: Libin Liu$^{3}$, \: Zilong Zhang$^{1}$, \: Zi Wang$^{1}$, \: Hongzhao Xie$^{2}$, \: Tengyu Liu$^{2}$, \: Xinyi Tong$^{2,4}$, \: \textbf{Yaodong Yang}$^{3}$, \: \textbf{Zhaofeng He}$^{1}$ }
        \\
{\parbox{\textwidth}{\centering $^1$Beijing University of Posts and Telecommunications \quad{}
         $^2$National Key Laboratory of General Artificial Intelligence, BIGAI \\
         $^3$Peking University\quad{}
         $^4$Central Conservatory of Music
                \\ \quad{} \\
\large \href{https://learnedunimax.github.io/}{https://learnedunimax.github.io/}
       }
}
}

\usepackage{amsmath}
\usepackage{amsfonts}
\usepackage{subcaption}
\usepackage{amssymb}
\begin{document}

\teaser{
\vspace{-15pt}
 \includegraphics[width=0.98\linewidth]{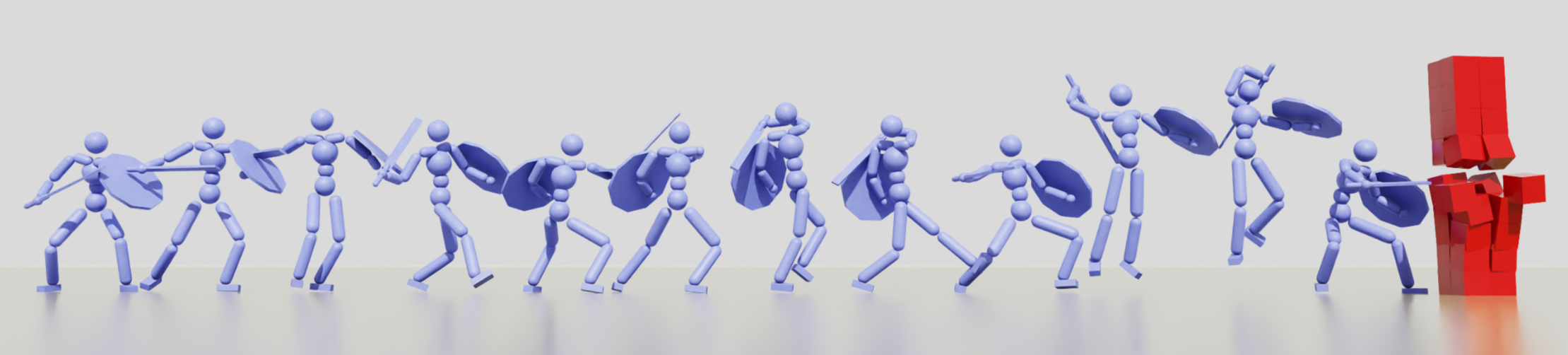}
 \centering
  \caption{We develop a versatile, adaptable, and controllable control model. Here, a character is depicted precisely performing an intended skill in an interactive control scenario.}
\label{fig:teaser}
}

\maketitle
\begin{abstract}
Learning natural and diverse behaviors from human motion datasets remains a significant challenge in physics-based character control. Existing conditional adversarial models often suffer from tight and biased embedding distributions where embeddings from the same motion are closely grouped in a small area and shorter motions occupy even less space. Our empirical observations indicate this limits the representational capacity and diversity under each skill. An ideal latent space should be maximally packed by all motion's embedding clusters. Although methods that employ separate embedding space for each motion mitigate this limitation to some extent, introducing a hybrid discrete-continuous embedding space imposes a huge exploration burden on the high-level policy. To address the above limitations, we propose a versatile skill-conditioned controller that learns diverse skills with expressive variations. Our approach leverages the Neural Collapse phenomenon, a natural outcome of the classification-based encoder, to uniformly distributed cluster centers. We additionally propose a novel Embedding Expansion technique to form stylistic embedding clusters for diverse skills that are uniformly distributed on a hypersphere, maximizing the representational area occupied by each skill and minimizing unmapped regions. This maximally packed and uniformly distributed embedding space ensures that embeddings within the same cluster generate behaviors conforming to the characteristics of the corresponding motion clips, yet exhibiting noticeable variations within
each cluster. Compared to existing methods, experimental results demonstrate that our controller not only generates high-quality, diverse motions covering the entire dataset but also achieves superior controllability, motion coverage, and diversity under each skill. Both qualitative and quantitative results confirm these traits, enabling our controller to be applied to a wide range of downstream tasks and serving as a cornerstone for diverse applications.


\end{abstract}  
\section{Introduction}

Synthesizing natural and lifelike character behaviors from human motion datasets has emerged as a significant research focus \cite{controlvae, won2022physics, tessler2023calm, peng2022ase}. An ideal control model of motor skills should support the generation of life-like motions with a high degree of diversity, reflecting the varied expressions and unique behavioral styles exhibited by humans across different skills. Another highly desirable property is that it should offer reusability, serving as a cornerstone for a range of downstream applications. A well-qualified reusable model should learn diverse, varied, and realistic motor skills, as it will function as a low-level control model directed by high-level strategies to complete tasks. Insufficient diversity limits the synthesis of natural behaviors. Furthermore, it should be capable of performing desired motions in response to specified skill commands, to support interactive control secnerios. However, developing such a versatile, adaptable, and controllable motor skill model remains challenging for the current state-of-the-art generative controllers.

\begin{figure}[htb]
\centering
\begin{subfigure}{0.50\linewidth}
\centering
\includegraphics[width=\linewidth]{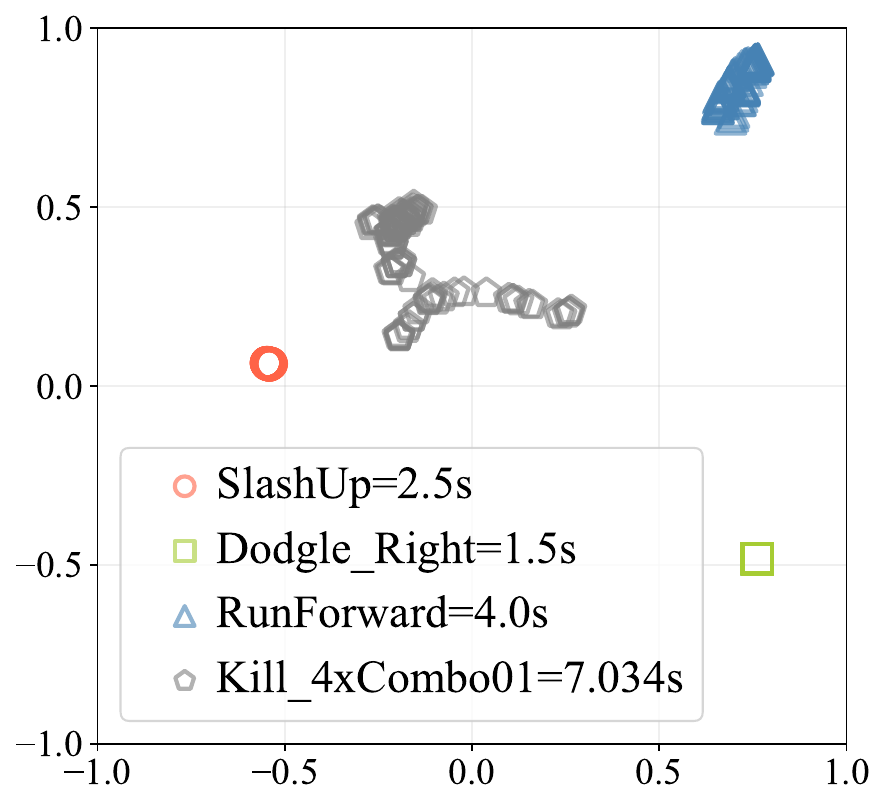}
\caption{CALM's latent space}
\label{Fig:distribution}
\end{subfigure}\hfill
\begin{subfigure}{0.50\linewidth}
\centering
\includegraphics[width=\linewidth]{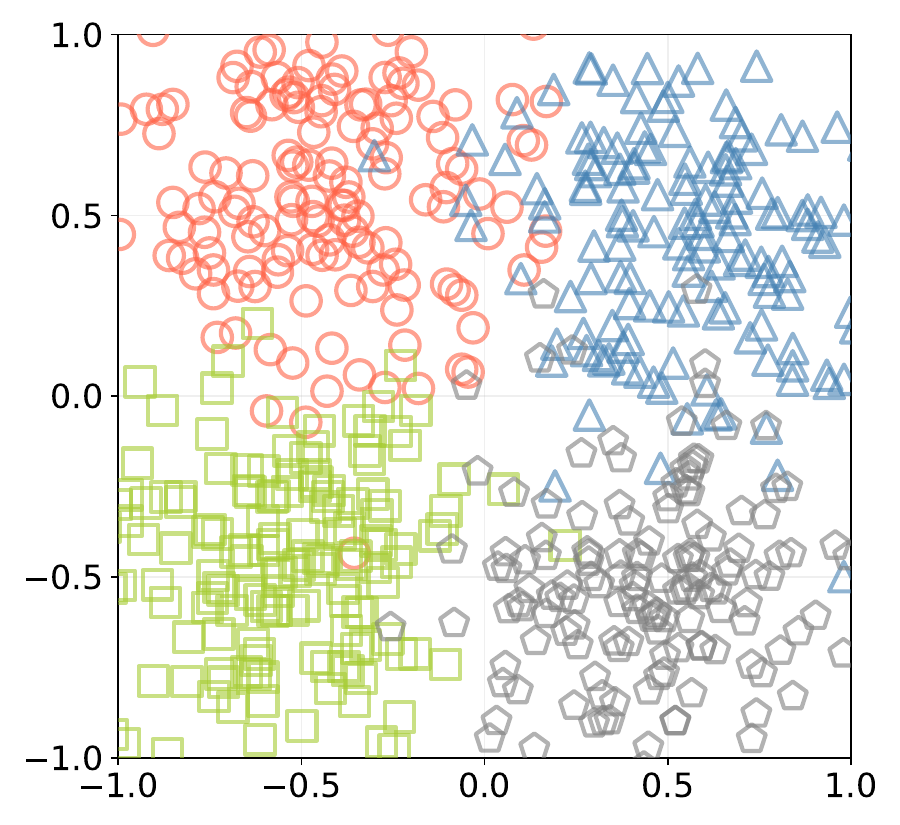}
\caption{Our latent space}
\label{Fig:humanoid_score}
\end{subfigure}
\caption{The visualization of two latent spaces using PCA. (a) The tight and biased embedding distributions for each motion in CALM, with labels indicating the motion names and their respective lengths. The spatial occupation of each skill correlates with the length of the motion clip. (b) The learned skill embedding clusters in our model form a maximally packed, uniform, and length-agnostic distribution.}
\label{Fig:reconstruc_score}
\end{figure}

Recent works have increasingly adopted Generative Adversarial Network \cite{goodfellow2020generative} (GAN)-based control model \cite{juravsky2022padl, dou2023c}. Some methods learn skill embeddings by projecting motions onto a unit hypersphere, which has been empirically shown to improve training stability, facilitate compact feature clustering, and ensure uniform class distribution \cite{chopra2005learning, xu2018spherical, wang2020understanding}. Compared to other methods that leverage explicit tracking rewards to precisely replicate each motion \cite{ncp, peng2018deepmimic}, GAN-based models can generate more diverse variations of a motion while also preserving its distributional behavioral characteristics. However, this advantage is impaired in current state-of-the-art conditional GAN-based control models due to the tight and biased embedding distribution on spherical latent space, as demonstrated in Figure~\ref{Fig:distribution}, which shows a part of the region in latent space from previous work \cite{tessler2023calm} after dimensionality reduction by Principal Component Analysis (PCA). These models \cite{tessler2023calm, juravsky2022padl} utilize segment-wise encoding to learn representations of reference motions, resulting in embeddings from the same motion being closely clustered on the surface of the spherical latent space. Notably, the spatial occupation of skill embeddings correlates with the length of the motion clips and shorter motions tend to occupy even less representational space. Our empirical observations indicate that this tight and biased clustering of skill embeddings restricts representational capacity, leading to reduced diversity for each skill. Furthermore, the embedding regions that remain unlearned could result in unnatural and meaningless movements. Both the restricted diversity of each skill and the existence of substantial unmapped regions adversely affect the performance of the controller. Methods like \cite{dou2023c} help alleviate the issue of limited diversity by constructing a discrete-continuous skill embedding space, allocating independent unit hyperspheres to each motion. However, as pointed out in \cite{li2021hyar}, the introduced heterogeneity in this hybrid low-level space poses significant exploration challenges. The high-level policy struggles to navigate and explore appropriate discrete skill labels while simultaneously approximating suitable continuous skill embeddings under the predicted label. This issue is particularly pronounced in tasks requiring the operation of various heterogeneous skills, such as in the strike task \cite{peng2022ase}, hindering the synthesis of natural and fluid strategies.

We hypothesize that an ideal skill representation should contain clusters of skills uniformly distributed over a latent embedding space and collectively covering the entire space. We refer to it as a maximally packed and uniformly distributed space. This 
ensures equal representational area for each skill in the latent space while reducing the presence of unmapped areas. However, achieving such a distribution on a spherical surface is a variation of the Tammes problem \cite{tammes1930origin}, which lacks straightforward solutions, necessitating a learning-based approach. Neural Collapse \cite{nc, galanti2021role, han2021neural} is an intriguing phenomenon observed in classification models, where the features of data points from the same class tend to converge toward their class mean, collapsing into a single point. Simultaneously, the class means themselves become equidistant and maximally-equiangular positioned in the feature space, thereby maximizing the inter-class separation and forming a simplex equiangular tight frame \cite{liu2023inducing, han2021neural}. This phenomenon potentially facilitates the modeling of uniform distributions for feature means of each motion clip in high-dimensional spherical latent space. 

\label{text:intro}

On the other hand, a naive implementation of neural collapse still fails to provide diversity within each motion, as the representation of each motion collapses to its feature mean for each skill. We propose an Embedding Expansion method to expand the areas occupied by each skill and maximally pack the skill clusters over the embedding space, maximizing the diversity potential under each skill. To this end, we develop a GAN-based skill-conditioned control policy that learns diverse skills with expressive variations from motion datasets. Through a classification-based encoder and embedding expansion, our model is compelled to learn all reference motions and construct embedding clusters for each motion. These skill-specific embedding clusters are uniformly distributed on a hypersphere with all clusters occupying equivalent areas. Skill embeddings within the same cluster generate behaviors that conform to the characteristics of the corresponding motion clips, while also exhibiting noticeable variations. Experiment results reveal that our controller is capable of generating diverse motions that cover the entire dataset, featuring high-quality and versatile variations. Compared to the existing methods, our method achieved superior controllability, motion coverage, and diversity under each skill. All these traits are further substantiated by its effectiveness in tackling complex downstream tasks and improving performance in interactive scenarios.

Overall, our contributions to this paper are as follows. 1) We exploit a classification-based encoder to establish a uniform distribution prior across motion clips, creating distinctive stylistic skill cluster centers. 2) Combined with conditional generative learning, we apply the Embedding Expansion to form embedding clusters for all skills with great variations. 3) Experiment results reveal that the above steps are well synergized, and collectively promote a versatile, adaptable, and controllable policy, becoming a cornerstone for varied downstream applications.

\section{Related Work}
\subsection{Physics-based
character control}
Early research in computer animation within physics-based environments \cite{yin2021discovering, yu2018learning, coros2010generalized} has frequently utilized various methods, including trajectory optimization \cite{muico2011composite, wampler2014generalizing, de2010feature, levine2013guided, mordatch2012discovery, mordatch2013animating}, equations of motion \cite{raibert1991animation}, policy search \cite{xie2020allsteps, tan2014learning}, and heuristic optimization \cite{grochow2004style, wang2009optimizing}. However, due to the absence of a reference motion, the character could produce unnatural behaviors. Data-driven methods address this issue by encouraging characters to imitate human motions However, due to the absence of a reference motion, the character could produce unnatural behaviors. Data-driven methods address this issue by encouraging characters to imitate human motions \cite{da2008simulation, ding2015learning,lee2010data,liu2012terrain, brown2013control}. Recent breakthroughs in deep reinforcement learning further facilitate the controller to learn more vivid behaviors by a tracking reward function \cite{peng2018deepmimic, fussell2021supertrack, bergamin2019drecon, won2019learning}, which encourages the model to mimic a sequence of reference pose states in a motion clip. Subsequent efforts  \cite{park2019learning, peng2019mcp, peng2021amp} aimed to integrate natural motor skill acquisition with life-like behavior task completion. However, the acquired motor skills lack reusability, requiring relearning of low-level controls even when two tasks share the same reference motion dataset. Thus, decoupling task-solving from natural behavioral learning and enabling motor skill reusability offers a more efficient approach to physics-based character control.

\subsection{Learning reusable priors for character control}
To enable the reusability of motor skills, some recent works \cite{merel2018neural, won2022physics, hasenclever2020comic} model a low-level skill space by distilling tracking-based policies into a latent space using variational autoencoder (VAE) \cite{kingma2013auto} structures. To avoid generating out-of-distribution actions, model-based methods \cite{won2022physics, controlvae} employ a world model to ensure the actions taken are consistent with anticipated future states. However, the trade-off problem between reconstructive accuracy and conformity with the prior distribution inherent in VAE-based methods leads to a compromise between motion quality and model reusability. To address this challenge, some studies \cite{ncp, yao2024moconvq} opt for a quantized variational autoencoder-based approach, using discrete latent representations organized into a codebook To learn a more comprehensive range of motor skills, some researchers \cite{luo2023universal} use progressive neural networks and policy distillation to learn reusable skills from large-scale motion databases.

Another desired latent space is the unit hypersphere, which is primarily exploited by the generative adversarial network (GAN) \cite{goodfellow2020generative}-based method in physics-based character control. Recent works \cite{peng2022ase, xu2023adaptnet} learn reusable motor skill embedding through adversarial motion priors \cite{peng2021amp}. They encode the skills in the dataset into stylistic latent onto their spherical latent space. However, the mode-collapse limits the motion diversity that fails to cover the entire dataset. To achieve a high motion coverage, some researchers have been developing conditional GAN-based models \cite{juravsky2022padl, tessler2023calm}, which compel the policy to generate motions coherent with the behavioral style specified by the conditioned reference motion. While these models offer interactive control potential, the generated motions by desired skill embeddings are still undetermistic. One reason is the stylistic embeddings learned by GAN are adaptable and often produce variant sub-motions of original motion clip. Although this facilitates improved synthesis of strategies for downstream tasks, for interactive control tasks that require precise responses to user commands, such unpredictable control can undermine the user experience. Additionally, the encoder in these methods applies segment-wise encoding to two-second subsequences \cite{tessler2023calm, juravsky2022padl}, leading to a limited diversity and biased distribution of skill embedding for all motions. This limits diversity under each skill, impairing performance in downstream tasks. Subsequent work \cite{dou2023c} allocates a unit hypersphere for each motion clip, avoiding the restricted representational space problem. However, as discussed in Section.\ref{text:intro}, the inherent coarse discrete-continuous action space makes it difficult for high-level policies to synthesize natural strategies in complex tasks.

\begin{figure*}
    \centering
    \includegraphics[width=0.95\linewidth]{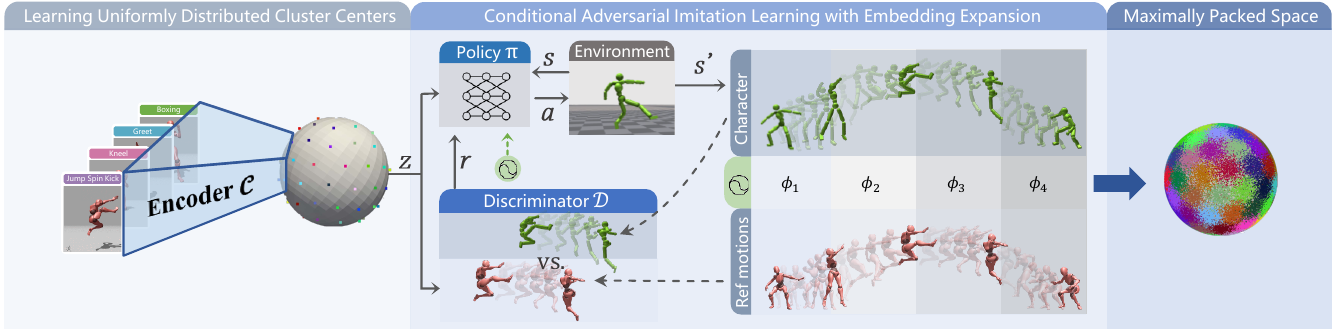}
    \caption{Our method uses a unit hypersphere as the embedding space to feature uniformly distributed embedding clusters for each skill. We first employ a classification-based encoder to distribute motion features uniformly on a high-dimensional sphere, then apply conditional imitation learning with the Embedding Expansion technique to form a stylistic skill embedding cluster for each skill, achieving a maximally packed and uniformly distributed space. }
    \label{Fig:Pipeline}
\end{figure*}

\section{Reinforcement Learning Background}
Controlling physically simulated humanoid characters is challenging. Our method employs reinforcement learning to develop a skill-conditioned controller, which
comprises state $s$, action $a$, transition probability $p$, reward $r$, and a discount factor gamma $\gamma$. At each timestep $t$, the control policy $\pi(a_{t}|s_{t}, z)$ calculates an action $a_t$ based on current state $s_t$ with a skill-specific embedding $z$ from the embedding space and applies $a_t$ on character. The black-box simulated environment returns the next state $s_{t+1}$ according to transition probabilities $\mathcal{P}$ with a gained reward $r$. The policy incrementally refines its ability to an impressive controller by maximizing the expected discounted return $J$,
\begin{equation}
J = \mathbb{E}_{p(\tau|\pi)} \left[\sum_{t = 0}^{\infty} \gamma^{\ t} r_t \right],
\end{equation}
where $p(\tau|\pi) = p(s_0) \prod_{t = 0}^{\infty} p(s_{t + 1} | s_t, a_t) \pi(a_t | s_t, z)$ is the probability over the set of all possible trajectories.
\section{Overview}
Our framework develops a skill-conditioned controller designed to extract and learn fluid and natural motions from a reference motion dataset into reusable, skill-specific embeddings.
Figure~\ref{Fig:Pipeline} provides a schematic overview of our system. We first learn a classifier-like encoder $\mathcal{C}$ to map motion clips in different skills into features. The feature means of each motion are evenly distributed across the spherical latent space $\mathcal{Z}$, laying the groundwork for the architecture of latent space. Due to the Neural Collapse phenomenon, the features of each motion clip tend to collapse to a single point, which is their feature mean and will serve as the cluster center for that specific skill. 

Building on this, we develop our control model using conditional imitation learning on embeddings from each skill, while jointly training a conditional discriminator $\mathcal{D}$ to guide the policy learning. These skill-specific embeddings are obtained by sampling using the Embedding Expansion on each feature mean. Through Embedding Expansion, the behavioral style of individual motion clips is learned into diverse stylistic embeddings around respective cluster centers. This constructs embedding clusters for each skill and demonstrates expressive in-skill diversity. During training, we embed motion-progress information into the character states exclusively when conditioning on each feature means. This selective integration introduces an additional control dimension to guarantee the execution of full movements, such as kinds of combos in video games, to clearly convey motion intentions.

After training, we obtain the learned control policy $\pi$, which can be applied to a variety of downstream motion synthesis applications. 1) The learned control policy can serve as a low-level controller in combination with learnable high-level control strategies $\omega(z|s, g)$ to accomplish complex downstream tasks with natural and fluid movements. During the task, the $\omega$ specifies the embedding $z$ to $\pi$ based on the task-specific goal $g$ and current task observation $s$. 2) Due to its controllability to precisely perform desired skills, the learned policy $\pi$ can be integrated with a locomotion high-level policy for interactive control scenarios.

\section{Learning Uniformly Distributed Cluster Centers}
In our work, stylistic embedding clusters corresponding to the respective reference motion are distributed uniformly in the latent space.
To achieve a balanced and uniform distribution of skill-related embedding clusters, it is essential to ensure that the centers of different clusters are evenly spread on the unit hypersphere, which is a case of Tammes problem \cite{tammes1930origin}. We approach this best-packing problem on the hypersphere by leveraging the Neural Collapse phenomenon observed in classification models. 
Given a dataset consisting of $n$ different motion clips $\mathcal{M}$, we learn a classifier-like encoder $\mathcal{C}$ by a classification task. In the dataset, each motion clip $m = \{{q}_t, 0 \leq t \leq L\} \in \mathcal{M}$ represents a sequence of character body states ${q}$ over a total length of L second. In our classification task, each motion clip is treated as a distinct class. We split each motion clip into overlapping sub-motions of 2 seconds in duration, serving as the training samples for that class. For motion clips that are shorter than 2 seconds, we use looping padding to extend them to the required length. The encoder $\mathcal{C}$ features an $l_2$ normalization module in its penultimate layer, which maps the features of each motion clip onto the surface of a unit hypersphere. These normalized features are then converted into probability vectors through a softmax layer. The learning of $\mathcal{C}$ is supervised by cross-entropy, aiming to minimize the difference between these predicted probability vectors with the true label distributions. As training progresses, the encoder $\mathcal{C}$ begins to approach the state of Neural Collapse, where intra-class variability significantly diminishes. The features extracted from each motion clip converge tightly to their class-specific mean representations, becoming nearly identical within each class. The feature means  $\mathcal{U}=\{u_1, u_2, ..., u_n|u_i=\mathcal{C}(m_i)\}$, where $\mathcal{U} \subset \mathcal{Z}$, become symmetrically arranged in the feature space. They form a simplex equiangular tight frame, where the angle between any pair of class means is almost equal. This arrangement promotes a uniform distribution of features and leads to a maximally separated and symmetric configuration on the hypersphere.

\section{Learning Skill-Specific Stylistic Embedding Clusters}
After the first stage, each motion clip has a corresponding feature mean, and these feature means are uniformly distributed on the surface of the unit hypersphere. In the subsequent phase, we employ Conditional Adversarial Imitation Learning, integrated with the Embedding Expansion technique, to develop our control model. During this stage, stylistic embedding clusters are formed with the feature means as their centers. Through Embedding Expansion, the model is enhanced to respond to the embeddings in the vicinity of the cluster center $u \in \mathcal{U}$, enabling it to produce a diverse range of movements. These movements are not only varied but also maintain stylistic consistency with the reference motion linked to their cluster center $u$. This ensures that the controller adapts flexibly to different motion contexts while preserving the essence of the original movements. 

\subsection{Conditional Adversarial Imitation Learning}
Our method adopts conditional adversarial imitation learning to learn skill-specific embeddings for each reference motion.
At each timestep, the control policy $\pi$ processes two inputs: the current character body state $s_t$ and a sampled embedding $z$ around the specified motion clip feature means $u$. The latter one will be elaborated in the Section.\ref{section:ee}, informing the controller about the specific skill to execute. The policy $\pi$ is learned under the supervision of the conditional discriminator $\mathcal{D}$. During this training stage, the discriminator critically evaluates state transitions $(s_t, s_{t+1})$ paired with the skill embedding $z$. The objective of the $\mathcal{D}$ is to evaluate whether the state transitions, in conjunction with the motion feature $z$, authentically represent the intended motion clip and verify that these transitions are derived from the authentic dataset rather than being generated by the policy $\pi$. Consistent with the prior conditional adversarial skill embedding learning works \cite{juravsky2022padl, tessler2023calm}, the loss function for our conditional discriminator is revised as:
\begin{equation}
\label{eq:disc}
\begin{aligned}
\mathcal{L}_D = & -\mathbb{E}_{p_{\mathcal{Y}}(\cdot)} \left[\log \left( D\left(s_t, s_{t+1}, z\right)\right)\right] \\
& -\mathbb{E}_{p_{\mathcal{N}}(\cdot)} \left[\log \left(1 - D\left(s_t, s_{t+1}, z\right)\right)\right] \\
& -\mathbb{E}_{p_{\pi}(\cdot)} \left[\log \left(1 - D\left(s_t, s_{t+1}, z\right)\right)\right] \\
& + w_{\mathrm{gp}} \mathbb{E}_{p_{\mathcal{Y}}(\cdot)} \left[\left\|\nabla_{\psi} D(\psi, z) \big|_{\psi=(s_t, s_{t+1})}\right\|^2\right],
\end{aligned}
\end{equation}
in which, the probability distribution \(p_{\mathcal{Y}}(\cdot)\) models scenarios where the transitions $({s}_t, {s}_{t+1})$ originated from motion $m$ and are correctly aligned with the embedding $z$. Specifically, $z$ belongs to the embedding cluster centered at $u = \mathcal{C}(m)$, ensuring consistency between the motion and its corresponding stylistic embedding. Conversely, the distribution \(p_{\mathcal{N}}(\cdot)\) represents mismatched scenarios. The state transitions $({s}_t, {s}_{t+1})$ are collected from $m$, but the embedding $z$ corresponds to a cluster whose center is the feature mean of a different motion, rather than $m$. This misalignment is used to train the discriminator to recognize and penalize incorrect pairings. The distribution \(p_{\pi}(\cdot)\) models the generation of the state transitions $({s}_t, {s}_{t+1})$ from the policy $\pi$ conditioned on a specified embedding $z$. To ensure stability in our adversarial training process, we incorporate a gradient penalty into the loss function, weighted by $w_{GP}$. The discriminator is trained along with the policy learning, providing implicit rewards for conditional motion imitation calculated as:
\begin{equation}
\label{eq:reward}
r_t = -\mathrm{log}\left(1 - D({s}_t, {s}_{t+1}, u) \right).
\end{equation}
\subsection{Embedding Expansion}
\label{section:ee}
In our method, each motion clip in the dataset represents a skill and conveys a unique behavioral style. Mastering the distinctive style and ensuring a high diversity under each style will offer more options for the high-level policy, contributing to the flexible synthesis of natural strategies for downstream tasks. Unlike previous approaches that pursue a tight embedding cluster in a compact representational space for each motion, our method aims to maximize the representational space for each motion and provide each skill with greater diversity potential. To achieve this, we propose the Embedding Expansion technique during conditional adversarial imitation learning.

During the training, distinct clusters of stylistic embeddings are formed around each motion's feature mean $u$. Clustering around a central point on the unit sphere can be modeled by a von Mises-Fisher (vMF) distribution. We apply the vMF to model the stylistic embedding distribution within each skill's embedding cluster. In our settings, each stylistic embedding cluster is modeled as $z_u \sim \text{vMF}({u}, \kappa)$, $z_u \in \mathcal{Z}$, where ${u}$ serves as the mean of each distribution. The concentration parameter $\kappa$ models the dispersion of data points around each center ${u}$ on the hypersphere space. The probability density function of each distribution determines the likelihood of sampling point $z$ around each cluster center, defined as:
\begin{equation}
f({z_u}; {u}, \kappa) = C_p(\kappa) \exp(\kappa {u}^{T} {z_u}) ,
\end{equation}
where the $C_p(k)$ is the normalization constant, ensuring that the total probability integrates to one. 

Through Embedding Expansion, the available embeddings of each skill, input to both the conditional discriminator and the policy, expand from original feature mean $u$ to $\{u, \text{vMF}({u}, \kappa)\}$. Each environment has a probability \( p \) of using the $u$, and a probability $(1-p)$ of substituting \( z \) in Equations~\ref{eq:disc} and~\ref{eq:reward} with a sampled embedding $z_u$. 
Thus, skill-specific stylistic embeddings are learned from different clusters. Because the centers of all clusters have been uniformly distributed across the latent space, which is determined by the classification-based encoder, all skill embedding clusters are evenly spaced in the embedding space. With the same concentration parameter $\kappa$ for all constructed vMF distributions, each cluster occupies an almost equal region of the embedding space, leading to balanced learning across all motion clips. Embeddings from the same cluster result in diverse motions, but all conform to the unique behavioral style depicted in the associated reference motion.

\subsection{Interval Motion-Progress Encoding for Cluster Center}

Most existing skill-conditioned controllers learn distinct behavioral styles from each reference motion. 
However, predicting subsequent motion from the past is inherently ill-posed since multiple possible decisions exist at each frame, even if we know the exact motion style. 
Moreover, since these GAN-based models are not required to replicate reference motions precisely but rather to flexibly produce stylized behavior, motions that contain a full sequence of meaningful movements such as varied combos often lose their value. We empirically observe that the controllers are more willing to generate a low-cost subsequence of the original motion clip to appease the discriminator with minimal risk. This unpredictable and relatively incomplete skill execution can significantly impair the user experience in interactive control tasks, such as in video games, where characters need to perform complex and visually appealing combo skills.

To this end, we employ interval motion-progress encoding as a trick to our adversarial imitation learning. The utilization of motion-progress information strategically preserves the temporal structure of each motion $m$. We apply this technique exclusively on the settings that feature mean $u = \mathcal{C}(m)$ are used as the specified skill embedding.
Specifically, We partition each motion clip into several distinct motion stages with a fixed duration of $l$ second. Therefore, for a motion clip with a total duration of $L$ second, the state at each time step $s_t \in m$ has corresponding motion-progress encoding according to: 
\begin{equation}
\phi(t) = PE\left(\left\lfloor t/l \right\rfloor\right) \quad \text{s.t. } t \leq L ,
\end{equation}
where $\lfloor \cdot \rfloor$ is the floor function and ${PE}$ is the positional encoding \cite{vaswani2017attention} utilizing periodic functions. This facilitates policy and discriminator in capturing the relative temporal relationship for consecutive motion phases. These encodings are dimensionally equivalent to state features, allowing them to be directly summed together as the progress-embedded character body state $\widetilde{s}_t=s_t + \phi(t)$ at time step $t$. More details of PE can be found in the supplements.

At each time step $t$, the control policy $\pi(a_t|\widetilde{s}_{t}, u)$ produces an appropriate action $a_t$ based on the current progress-embedded state $\widetilde{s}_{t}$ with a desired motion clip unique feature $u$. The policy tries to fool the conditional discriminator through the generated state transitions $(s_t, s_{t+1})$ under embedding $u$, embedded with motion-progress encoding $({\phi}(t), {\phi}({t+1}))$ collected at each time step. The discriminator will also be tasked with learning the progress-aligned motions, wherein the original state transitions $(s_t, s_{t+1})$ are replaced by progress-embedded state transitions $(\widetilde{s}_t, \widetilde{s}_{t+1})$ in Equations~\ref{eq:disc} and Equations~\ref{eq:reward}. For a detailed explanation of Equation~\ref{eq:disc} regarding the study of motion-progress aligned motions, we have included it in the supplementary materials. This design facilitates a full sequence motion generation by using this temporal encoding as a motion progress indicator. For time steps $t \geq L$ in the simulation, the motion-progress encoding will leave to \textit{none}, then the progress-embedded state reverts to plain state $s_t$. Note that the per-state progress encoding design allows for the inclusion of transitional steps that contain different progress encodings within the $(s_t, s_{t+1})$ fed into the conditional discriminator, which is crucial for preventing abrupt or jerky movements during interval stage transitions. 

During the inference, it is only necessary to provide the feature $u=\mathcal{C}(m)$ of the desired motion $m$, along with the progress-embedded state that is synchronized with the timing of the $m$, the policy is able to perform the desired motion from its initiation through its progression to its completion, exactly as the reference motion.

\section{Downstream Applications}
Now, we obtain a low-level control model with diverse skill embeddings held in a spherical latent space decoder. To further enable the controller to facilitate varied downstream applications in natural behaviors, we construct several high-level policies for different tasks.

\subsection{High-level tasks}
The high-level policy, denoted as $\omega(z|s, g)$, is designed to operate low-level skills from the latent space $\mathcal{Z}$ to direct the low-level policy $\pi$, with the ultimate goal of achieving a task objective $g$. Similar to the previous \cite{peng2022ase, tessler2023calm, dou2023c}, the high-level policy $\omega$ first generates an unnormalized latent variable $\overline{z}$, which is then mapped onto the unit hypersphere by applying L2 normalization, resulting in $z = \overline{z} / ||\overline{z}||$. This normalized latent variable $z$ is subsequently decoded by the low-level policy $\pi$ into concrete actions. This structure efficiently bridges high-level objectives with low-level action control, enabling the synthesis of strategies for different tasks.

Beyond synthesizing motion across different skills, the high-level policy $\omega$ can also handle tasks with specific stylistic constraints. Our controller learns an embedding cluster for each motion, capturing a diverse set of stylistic skill embeddings. By introducing a reward function that encourages the use of embeddings from a specified cluster, we can leverage the cluster center as prior knowledge to guide $\omega$ in executing tasks with a desired behavioral style. 
The reward for the $\omega$ is then designed as:
\begin{equation}
r_t = w_G\text{ }r^G(s_{t}, a_t, s_{{t+1}}, g) + w_S\text{ }r^S(u) ,
\label{eq:style}
\end{equation}
in which \(r^G\) and \(w_G\) represent the reward function and the weight for the task reward. The term \(w_S\) denotes the weight for the style-constraint control reward. We use $r^S(u)=(u^Tz)$, the $u$ is the center of the embedding cluster corresponding to the specified behavioral style in motion $m$, obtained by $\mathcal{C}(m)$, and \(z\) is the high-level action generated by \(\omega\). For tasks that require the synthesis of multiple distinct skills, like the Strike task \cite{peng2022ase}, only the $r^G$ is used to guide task completion. For tasks required with specific stylistic constraints, such as certain locomotion tasks, the $r_S$ is incorporated to incentivize adherence to the desired style.

\subsection{Ineractive control tasks}
Our controller is capable of executing the intended skills, making it well-suited for interactive control scenarios. Leveraging the interval motion-progress encoding technique, the controller provides a more precise and predictable control interface. We propose a synergistic framework that combines the strengths of both a low-level policy $\pi$ and a high-level policy $\omega$. The $\omega$ is trained on a locomotion task, acting as a direction controller and responding to the input directional control signals. Simultaneously, users can invoke specific skills on-demand by directly interfacing $\pi$, enhancing the character's playability and responsiveness.

\begin{figure*}
    \centering
    \begin{subfigure}{\linewidth}
        \centering
        \includegraphics[width=\linewidth]{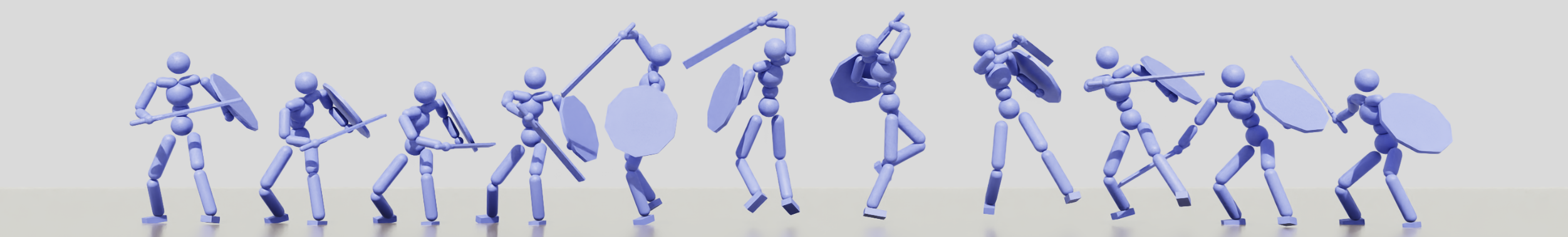}
        \caption{Warrior character on \textit{Sword\&Shield} dataset}
        \label{Fig:sub1}
    \end{subfigure}
    
    \vspace{1em} %

    \begin{subfigure}{\linewidth}
        \centering
        \includegraphics[width=\linewidth]{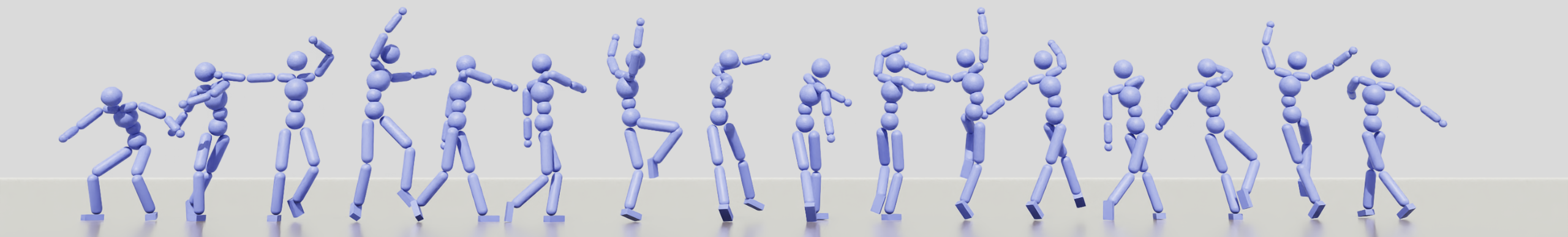}
        \caption{Humanoid character on \textit{MSA} dataset}
        \label{Fig:sub2}
    \end{subfigure}
    
    \caption{Our controller is capable of generating smooth, natural motions.}
    \label{Fig:demo_fig}
\end{figure*}

\section{Experimental Results}
We conducted comprehensive evaluations of our method, focusing on three key aspects: 1) the quality and data coverage of the motions generated by our policy on embeddings across the entire spherical latent space, 2) the motion diversity under each skill, and 3) the motion completeness of each desired skill in interactive control. Our evaluations involved two characters: a humanoid character with 28 degrees of freedom and a warrior character equipped with a sword and shield featuring 31 degrees of freedom. The humanoid character's control policy was trained using the \textit{MSA} dataset, which includes 123 various distinct motions with an average motion length of approximately 4.1 seconds at 30 FPS. It covers various human behaviors from locomotion skills, dance movements, and sports motions to acknowledged challenging or acrobatic skills. These motion clips are primarily sourced from Mixamo and the CMU Motion Capture Database. For the warrior character, we trained the control policy on the \textit{Sword\&Shield} dataset, open-sourced by \cite{peng2022ase, tessler2023calm}, which comprises 87 motion clips of different combos and few locomotion skills, with an average motion length of approximately 6.2 seconds at 30 FPS. The details of the dataset and designing of the state and action are elaborated in supplementary materials.

Our simulation environments are built using Isaac Gym \cite{makoviychuk2021isaac}. All experiments were conducted on a single NVIDIA GTX 4090, simulating 3072 environments in parallel. The simulation operates at 120Hz, employing a PD controller for character control, while the control policy is executed at 30Hz. The classification-based encoder is optimized by Adam with a learning rate of 0.01 for a total of 2000 epochs. Our policy is trained using Proximal Policy Optimization (PPO) \cite{schulman2017proximal} about 6 billion samples until coverage. Detailed hyperparameters of our encoder, policy, and discriminator are listed in the supplements. 

\subsection{Low-level policy}
Our framework learns a versatile low-level control model with a well-structured latent space. Different clusters are uniformly distributed on the surface of the unit hypersphere, in which embedding associated with the same reference motion are clustered around their respective feature means. We examine the efficacy of the learned control policy with its spherical latent space from the following perspectives.

\textbf{\textit{Motion Quality and Dataset Coverage.}} To showcase the motion quality and dataset coverage of the method, we compare our method against the state-of-the-art approaches CALM \cite{tessler2023calm} and c-ASE \cite{dou2023c}, both of which are leading conditional GAN-based models. Notably, both our method and CALM utilize a unit spherical space to host all motion skill embeddings. In contrast, c-ASE allocates a separate spherical space for the skill embeddings of each motion, requiring the policy to utilize both skill embedding information and an additional 64-dimensional skill label to identify the current skill being executed. For the warrior character, we train the CALM and c-ASE on the \textit{Sword\&Shield} dataset. For the humanoid character, we train these models on the \textit{MSA} dataset. All models are trained using their official implementations until the reward converges before being evaluated.

To ensure a fair comparison, we randomly sample 512 skill embeddings from the encoded representations of all motion clips in the CALM model. Additionally, to evaluate the impact of mapped region sparsity on model performance, we also conduct another experiment by assessing CALM using 512 randomly sampled latent variables from the entire hypersphere. For c-ASE, we follow the approach described in their paper, which involves using only one clip per skill label and generating 512 trajectories based on random skill labels and latent codes. 
For our model, we collect 512 skill embeddings that are randomly sampled from the latent space.
The length of collected motions for all models is the same, with over 600 steps. We evaluate the motion quality and diversity using the reconstruction score defined in \cite{ncp}, comparing the generated motions against reference motion frames by
\begin{equation}
\label{eq:recon}
s(f_i^*) = \max_{f \in \mathcal{F}} \left( 0.5 \cdot r^{jp}(f_i^*, f) + 0.5 \cdot r^{v}(f_i^*, f) \right),
\end{equation}

where the $f \in \mathcal{F}$ is the frame in generated motions and $f_i^*$ represent the frames in reference motions. The reconstruction score is measured through the similarity between $f$ and $f_i^*$ in  joint positions and root velocity, defined as $r^{jp}(f_i^*, f)$ and $r^{v}(f_i^*, f)$ respectively.

\begin{figure*}[htb]
\centering
\begin{subfigure}{0.50\linewidth}
\centering
\includegraphics[width=\linewidth]{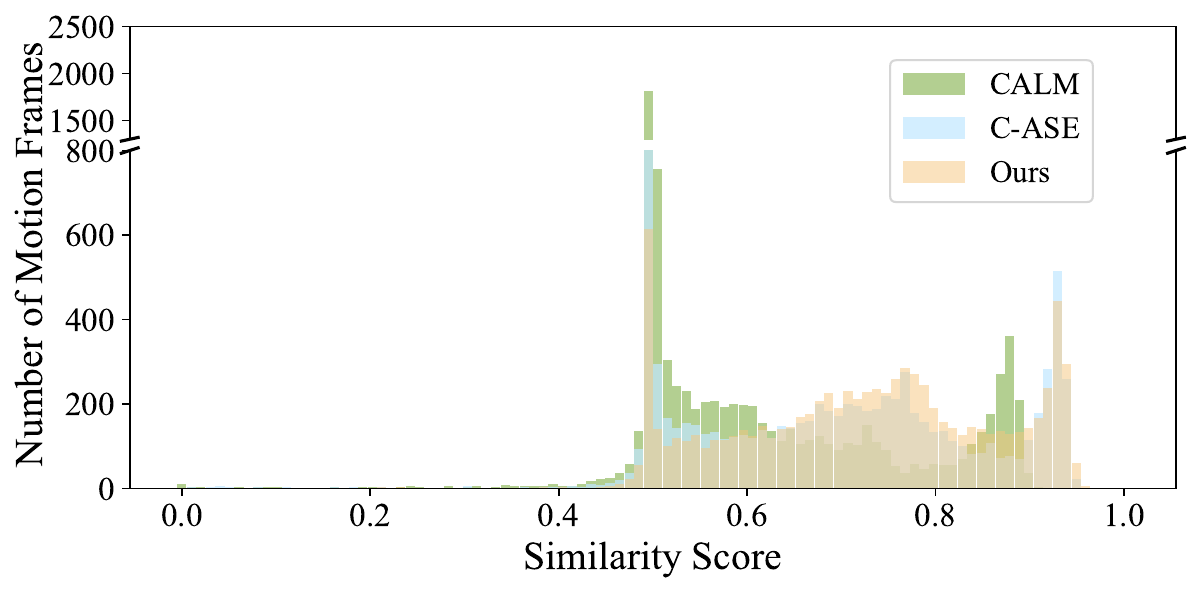}
\caption{Warrior character on \textit{Sword\&Shield} dataset}
\label{Fig:sword_score}
\end{subfigure}\hfill
\begin{subfigure}{0.50\linewidth}
\centering
\includegraphics[width=\linewidth]{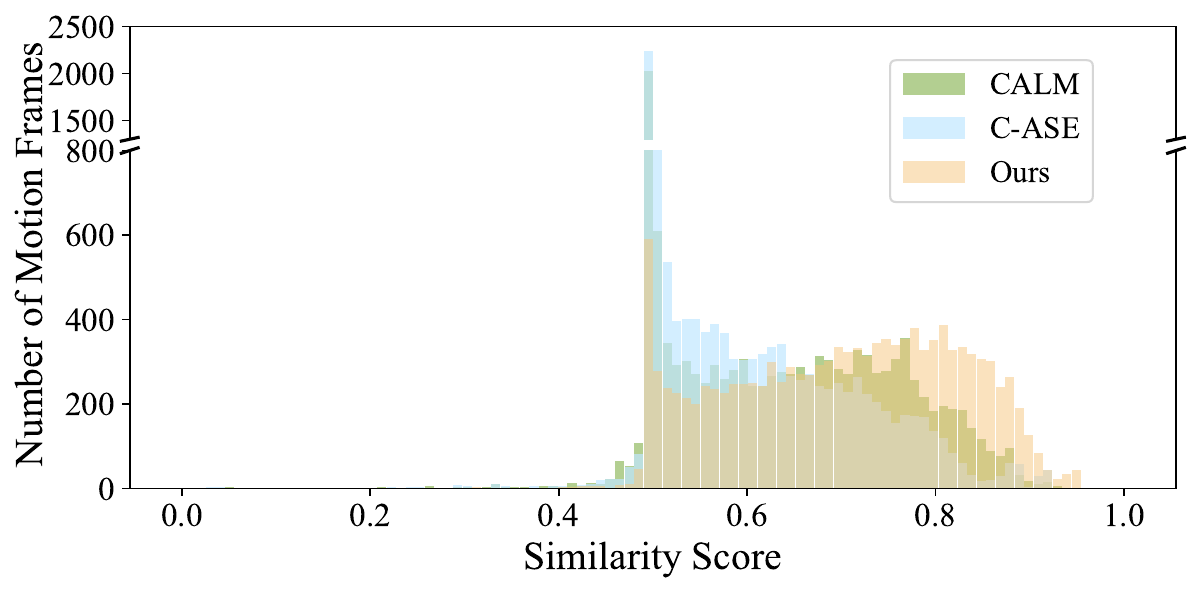}
\caption{Humanoid character on \textit{MSA} dataset}
\label{Fig:humanoid_score}
\end{subfigure}
\caption{The reconstruction scores achieved by CALM, c-ASE, and ours for the \textit{Sword\&Shield} dataset and  \textit{MSA} dataset. The horizontal axis represents the calculated reconstruction scores, while the vertical axis indicates the number of frames that achieved these scores.}
\label{Fig:reconstruc score}
\end{figure*}

Figure~\ref{Fig:sword_score} displays the reconstruction scores of the three models for the warrior character when comparing the frames in the dataset to the most similar frames in generated motions. The \textit{Sword\&Shield} dataset is characterized by a large number of motions with similar start and end poses, making it easier for the models to capture the overall behavioral style depicted in the dataset. A greater area on the right side of the histogram indicates a significant portion of the frames are reconstructed with low error, suggesting that the controller achieves better performance in terms of both motion quality and dataset coverage. As shown in Figure~\ref{Fig:demo_fig}, for \textit{Sword\&Shield} dataset, our model's latent space generates more natural and realistic motions, sufficiently covering almost the entire dataset. Although the distribution of the results in the bar chart for c-ASE is comparable to ours, a significant portion of the frames in the dataset still show relatively low reconstruction scores, ranging from 0.4 to 0.6. The majority of frames reconstructed by the CALM model received suboptimal scores. We hypothesize that limited representation capacity prevents the model from capturing full motion sequences, focusing instead on sub-sequences. Additionally, for the CALM model, the stylistic similarity of motions in the \textit{Sword\&Shield} dataset hampers its encoder's ability to differentiate motions, leading to averaged features and mediocre reconstruction performance. Quantitatively, our model achieved a reconstruction score above 0.5 for 91.2\% of the frames across all motion clips, with 73.5\% for CALM, and 82.3\% for c-ASE.

Figure~\ref{Fig:humanoid_score} displays the performance results on the \textit{MSA} dataset. Unlike the \textit{Sword\&Shield} dataset, which features motions with similar poses and consistent motion styles, the \textit{MSA} dataset presents a broader and more diverse set of behaviors. This significantly challenges the model’s learning capabilities, as it has to capture the distinct style of each motion. In this case, our model achieved a reconstruction score above 0.5 for 94.3\% of the frames, significantly outperforming CALM and c-ASE, which scored 80.2\% and 79.3\%, respectively. In the context of c-ASE, the increase in the number of skill labels poses a scalability challenge to their policy management. The extensive embeddings within each label hosted in the hyper-action space leads to many skills receiving plain reconstruction scores. For the CALM, the model exhibits better performance on these heterogeneous motions. 

For the CALM model employing random latents sampled from a hypersphere, there is a notable decline in the reconstruction scores—falling from 73.5\% to 64.7\% (a 12.0\% decrease) of the frames in the \textit{Shield\&Sword} dataset, and from 80.2\% to 57.1\% (a 28.4\% decrease) in the \textit{MSA} dataset. Visual results suggest that these declines are associated with large unmapped regions within the CALM latent space, which lead to unnatural movements, such as struggling to maintain alert postures. More visual comparison examples are shown in the video materials to demonstrate the effectiveness of our model in both motion quality and dataset coverage.

\textbf{\textit{Motion Diversity Under Each Skill.}} We demonstrate the diversity of stylistic embeddings for each motion clip's corresponding cluster in our method by comparing with the conditional model CALM and c-ASE. We select two intended motions from \textit{Sword\&Shield}: RunForward, a 4-second motion, and DodgeRight, a 1.5-second motion. For the CALM model, we randomly sampled 256 embeddings from the encoded motion features of each specified motion, denoted as $z \sim {E}_{\text{calm}}(m)$. We manually set the skill label embedding of $m$ for c-ASE and the 256 skill embeddings are randomly sampled from the respective hypersphere $\mathcal{Z}_c(m)$. Our model uses vMF distributions to model each motion's stylistic embedding cluster. In this experiment, we use the feature mean for $m$ as the skill embedding $z = \mathcal{C}(m)$. Meanwhile, we test 256 embeddings sampled from the vMF distribution of intended motion, $z \sim \text{vMF}(\mathcal{C}(m), \kappa)$. The value of $\kappa=50$ remains consistent with the training phase. The collected embeddings are then input into respective learned models, and each model controls 256 warrior characters with the same initial state for 500 timesteps.

\begin{figure*}[htb]
\centering
\begin{subfigure}{0.50\textwidth}
\centering
\includegraphics[width=\linewidth]{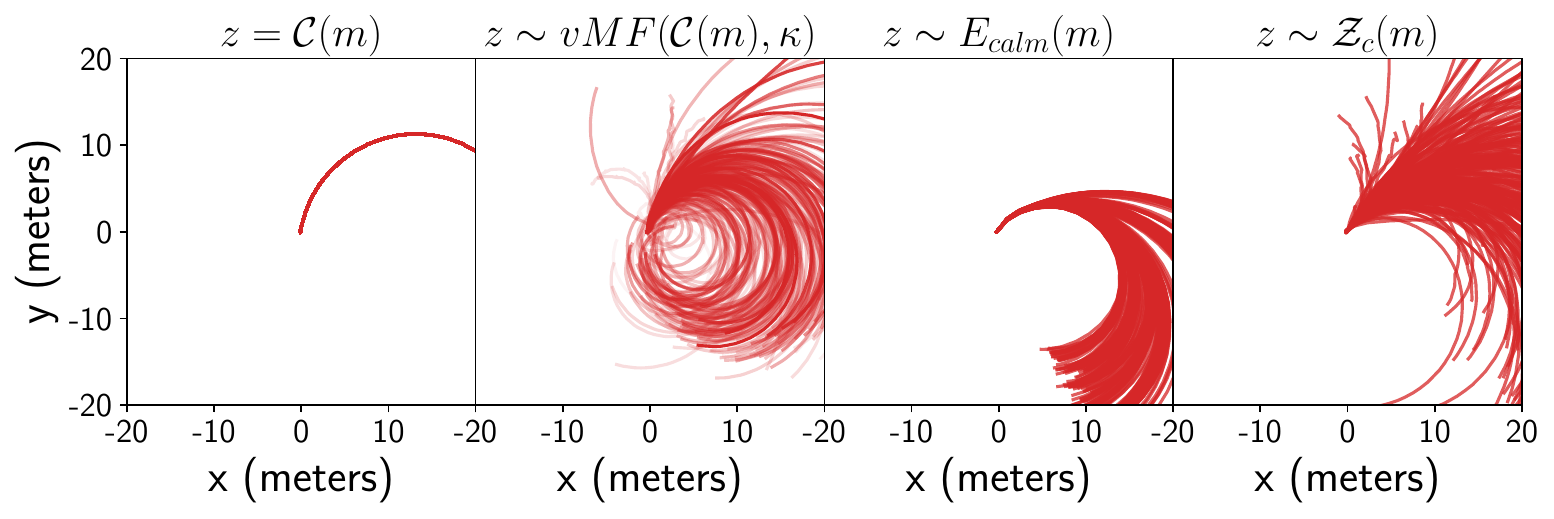}
\caption{Trajectories of the 256 characters' root on RunForward motion.}
\label{Fig:41}
\end{subfigure}\hfill
\begin{subfigure}{0.50\textwidth}
\centering
\includegraphics[width=\linewidth]{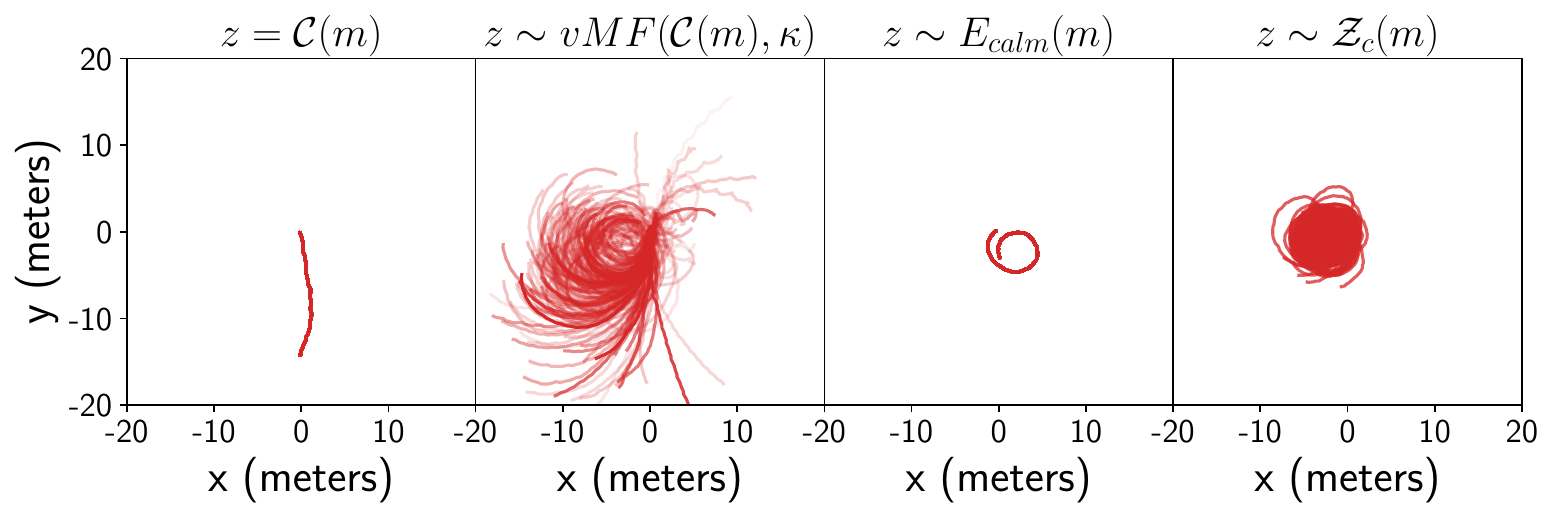}
\caption{Trajectories of the 256 characters' root on DodgeRight motion.}
\label{Fig:33}
\end{subfigure}
\caption{Trajectories of the characters' root position collected from CALM and our model. Plots in each sub-figure, from left to right, are our model conditioned on mapped motion's feature mean, our model conditioned on embeddings from an embedding cluster of associated motion, the CALM model conditioned on learned motion embeddings,  and the c-ASE model conditioned on desired skill label and random skill embedding.}
\label{Fig:location}
\end{figure*}

Figure~\ref{Fig:41} illustrates the trajectories of RunForward and DodgeRight motions generated by our model, CALM, and c-ASE using learned embeddings. The results demonstrate that our model produces a wide dispersion of trajectories, indicating our model effectively captures the length-agnostic behavioral style of each motion clip and produces a rich variety of movements. Notably, the trajectories generated by the cluster center of our model closely align with the reference motion, achieving a high level of fidelity. The in-skill diversity achieved by our model is further highlighted in the supplementary video materials, where nearly all characters exhibit natural and life-like running behavior with noticeable variations in movement speed and direction. In contrast, the CALM model shows a constrained diversity within the same skill, this limitation becomes more pronounced in shorter-duration DodgeRight motions. Across 256 characters, there is virtually no variation in the DodgeRight motions generated by CALM, underscoring its reduced capacity for diverse motion generation compared with ours and c-ASE.

\textbf{\textit{Motion Completeness.}} One key advantage of conditional models over other control models is their controllability to precisely execute specified skills. Our model enhances this capability by using motion-progress embedded states. This enables the execution of a full motion. This property is beneficial for interactive control scenarios like video games to ensure complete skill execution. The majority of motions in the \textit{Sword\&Shield} dataset consist of different kinds of combos, where the total moves in each combo range from a single rapid slash to a fluid sequence of five successive attacks. We use these attack motions as different skills and conduct comparative experiments on motion completeness during desired skill execution against the SOTA conditional control models. For the CALM, the reference motion is mapped by the encoder into embeddings, which are then used to control the generation of desired skills. we randomly select five embeddings from the encoder, and each embedding is executed for 300 steps, yielding a total of 1500 steps per skill. For c-ASE, we specify the intended skill label for policy and randomly sample 5 embeddings from its latent space, executing each for 300 steps. In contrast, our model uses only the feature mean $u$ of each skill as the conditional skill embedding. Unlike other methods where the policy directly conditions on the current states, our policy conditions on the progress-embedded state based on the current playtime, executing each combo skill over 300 time steps. If the playtime exceeds the total duration of the corresponding skill's motion clip, the motion-progress encoding become a void value and the progress-embedded states revert to the normal states. It is important to note that the progress-embedded states are only effective when the skill embedding is set to a feature mean from $U$. 

\begin{figure}[b!]
    \centering
    \includegraphics[width=0.95\linewidth]{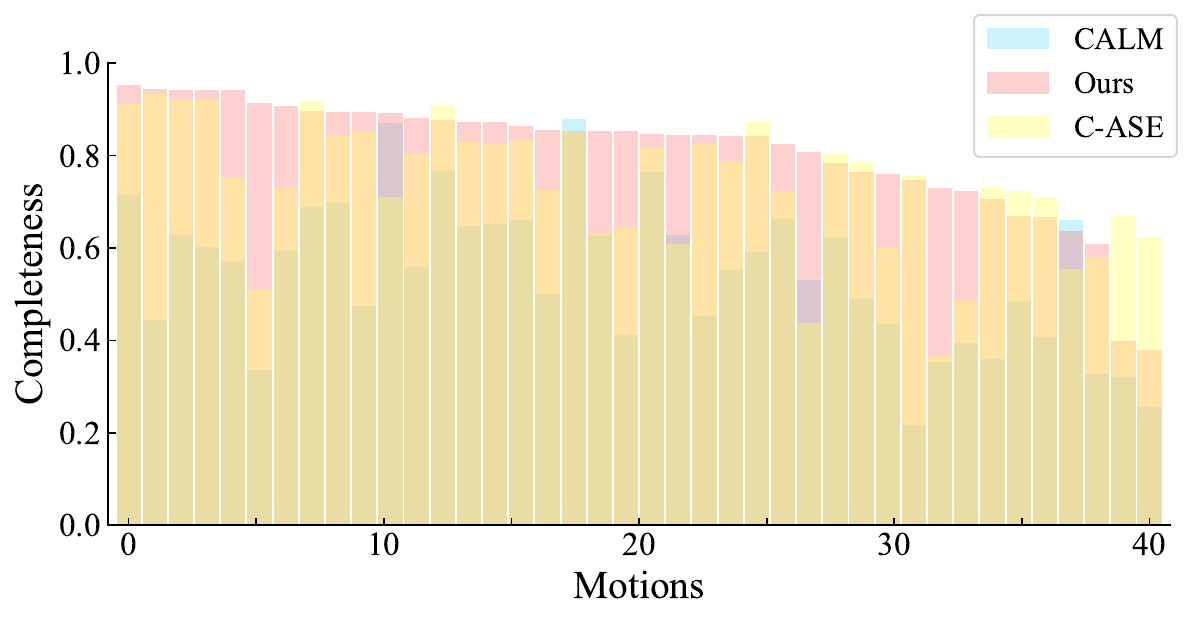}
    \caption{Motion completeness comparison between our motions generated by CALM, c-ASE, and ours on combo move skills.}
    \label{fig:completes}
\end{figure}

We then compare the generated motions of each skill to the frames in the corresponding reference motion. Using the construction score calculation method used in Equation~\ref{eq:recon}, we identify the reference motion frame closest to each generated frame and mark it as a covered frame. The motion completeness is then assessed by the ratio of covered frames to the total frames for each skill, depicted in  Figure~\ref{fig:completes}. The horizontal axis represents the evaluated combos, and the vertical axis represents completeness. By incorporating the motion-progress encoding as a progress indicator, our model achieves superior completeness across nearly all combo skills compared to CALM and c-ASE. Our model attains an average completion rate of 80.7\%, a 52.7\% improvement over CALM's average of 53.1\%, and a 9.5\% improvement over c-ASE's average of 73.7\%. Examples in the supplementary video will demonstrate this capability more vividly.

\subsection{Downstream Tasks}
The ultimate goal of designing and developing the control model is to apply it to a wide range of animated content generation tasks. We conducted several experiments across different downstream tasks to evaluate and analyze how different low-level policy architectures influence the downstream task.

\textbf{\textit{High-Level Tasks.}}
We conducted two high-level tasks with both humanoid and warrior characters: the location task, and the strike task. In the location task, the character is required to reach a specified target location using a specific behavioral style. This task is designed to evaluate the control model on the learned stylistic embeddings of each skill and investigate the impact of in-skill diversity on downstream tasks. For the strike task, the characters must approach a target object and knock it down, which intuitively requires the high-level policy to synthesize motions from heterogeneous skills into a natural strategy. This further tests the versatility and reusability of the control model. More details about the design of each task are provided in the supplementary materials.

We evaluated CALM, c-ASE, and our model on location tasks with RunForward and DodgeRight as style constraints. All models are treated as the inference-only low-level controller and utilized by a high-level policy trained to accomplish the task. The high-level policies of CALM and ours are trained to maximize both task reward $r^G$ and style reward $r^S$. For c-ASE, in which each skill is encoded in a separate latent space, the skill label was fixed to the corresponding label embedding and only $r^G$ is used during task training. Thus, all of the high-level policies only needed to estimate the skill embedding for the low-level controller. Our results show that all models successfully completed the task when RunForward was imposed as the style constraint. However, under the DodgeRight style, CALM failed to perform the task in the desired manner due to insufficient diversity under DodgeRight skill. In contrast, our model, which leverages uniformly distributed embedding clusters to provide greater expressiveness and variation within each skill, was able to perform the task in the desired style. Similarly, c-ASE also completes the task for the large representational space for each skill. More visual results are provided in the supplementary video.

In our experiments on strike tasks, which demand a more diverse combination of different skills, the high-level policies of CALM and our model are learned without considering the style reward $r^S$, operating freely within the low-level skill space. For the c-ASE model, we followed the implementation details outlined in their paper, learning a high-level policy that outputs both a skill label and a skill embedding simultaneously. The high-level policy produces a continuous probability distribution over skill labels, which are then converted into discrete skill labels using the Gumbel-Softmax, and a latent variable is sampled from a spherical Gaussian distribution. Our model achieves the task in a more lifelike manner: the character first engages locomotion skills to approach the target, then naturally transitions to aggressive skills to forcefully knock down the object. The CALM model also completed the task but often employed low-dynamics motions, such as wrist rotations, to topple the object. For c-ASE, we observed that their high-level policy struggles to maximize the task reward. It exhibited frequent switching between skill labels, leading to noticeable jitter and unnatural behavior in the character's actions. We hypothesize that this challenge stems from the hybrid discrete-continuous action space, which imposes a significant exploration burden on the high-level policy. Additionally, since the embedding space of each motion clip in the dataset is independent, it is
difficult for the control model to synthesize plausible transitions between different skills. Further experiments indicate that dividing heterogeneous skill motions into homogeneous sets to learn unified skill embeddings, for example, treating combos and locomotion behaviors as the same skill, alleviates this issue to some extent. However, this approach causes the c-ASE to regress into a behavior similar to that of ASE \cite{peng2022ase}, suffering from mode collapse. This contradicts the original design intent of c-ASE and results in the controller sacrificing controllability. 

\textbf{\textit{Ineractive Control Tasks.}}
To validate the performance of our model in interactive character animation, we designed a task similar to the previous approach \cite{tessler2023calm}. During the task, the user is required to interactively control the warrior character to knock down objects in the environment. We developed a synergistic framework where a high-level location policy governs character navigation, while the low-level policy not only executes the skill embeddings from the high-level policy but also carries out the motions commanded by the user. In our setting, the execution of user-commanded motions takes precedence over high-level policy control. The user can navigate the character to approach the target, then perform a desired motion $m$ by conditioning the low-level control on the motion’s feature mean $u = \mathcal{C}(m)$. During command motion execution, the policy uses progress-embedded state control to carry out the specified action as a full sequence. Compared to other controllable methods \cite{tessler2023calm, dou2023c}, a significant advantage of our approach is that by utilizing the interval motion progress information, our controller can generate more deterministic and predictable desired motions. This allows users to precisely control the start and end of a skill, resulting in a better interactive experience. The example of the task is demonstrated more vividly in the supplementary video.

\begin{figure}[ht]
    \centering
    \includegraphics[width=0.97\linewidth]{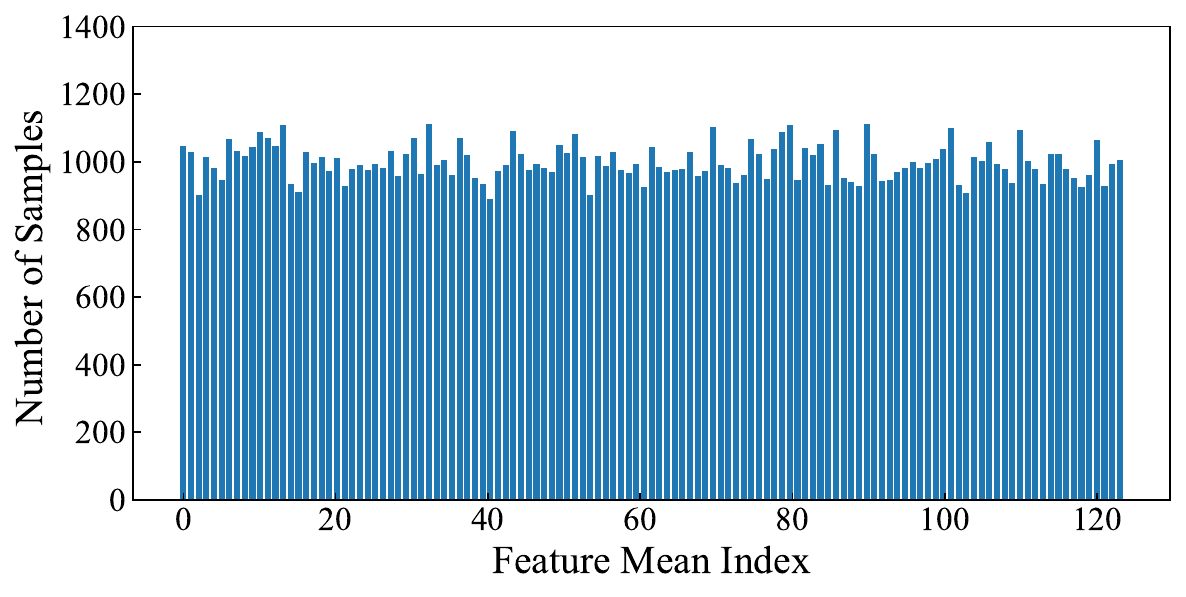}
    \caption{The number of embeddings located in the vicinity of each feature means.}
    \label{Fig:nc}
\end{figure}

\begin{figure*}[htb]
\centering
\begin{subfigure}{0.50\linewidth}
\centering
\includegraphics[width=\linewidth]{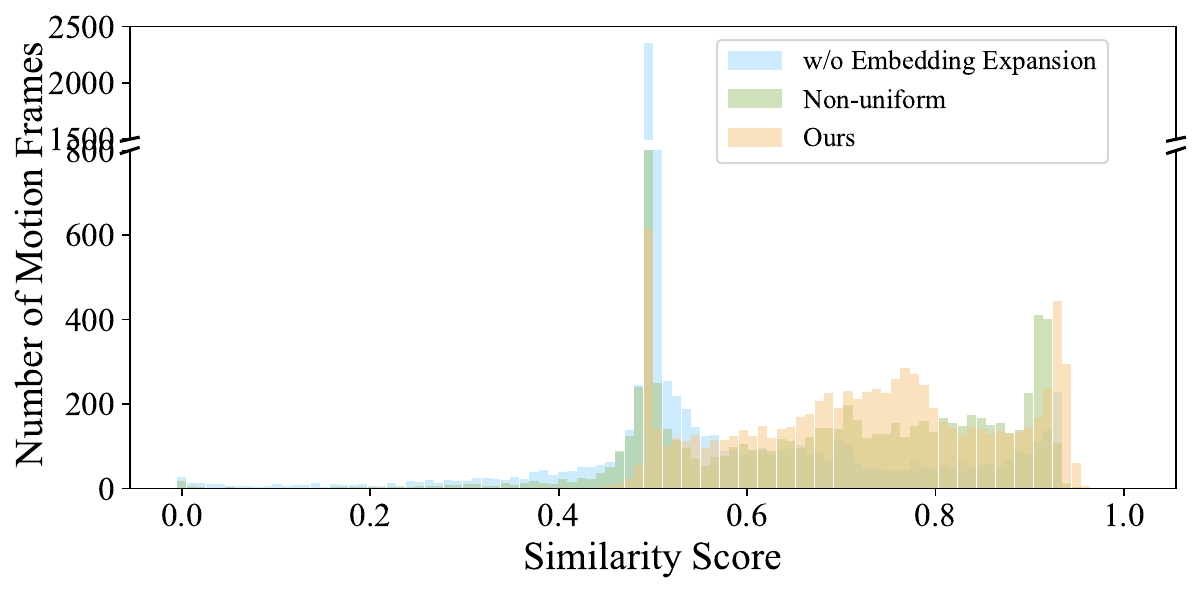}
\caption{Warrior character on \textit{Sword\&Shield} dataset}
\end{subfigure}\hfill
\begin{subfigure}{0.50\linewidth}
\centering
\includegraphics[width=\linewidth]{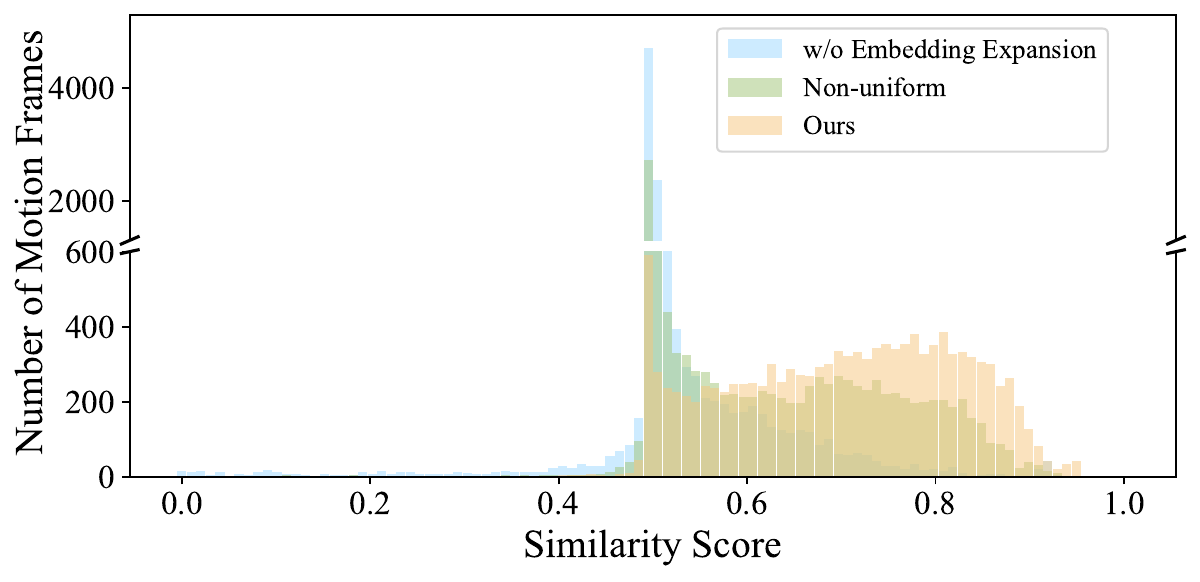}
\caption{Humanoid character on \textit{MSA} dataset}
\end{subfigure}
\caption{The reconstruction scores are achieved in three models for the \textit{Sword\&Shield} and \textit{MSA} dataset: our model without applying Embedding Expansion, our model without achieving uniform distribution, and a complete model. The horizontal axis represents the calculated reconstruction scores, while the vertical axis indicates the number of frames that achieved these scores. }
\label{Fig:reconstruc}
\end{figure*}

\subsection{The Effectiveness of Classification-Based Encoder}
\begin{figure}[htb]
\centering
\begin{subfigure}{0.50\linewidth}
\centering
\includegraphics[width=\linewidth]{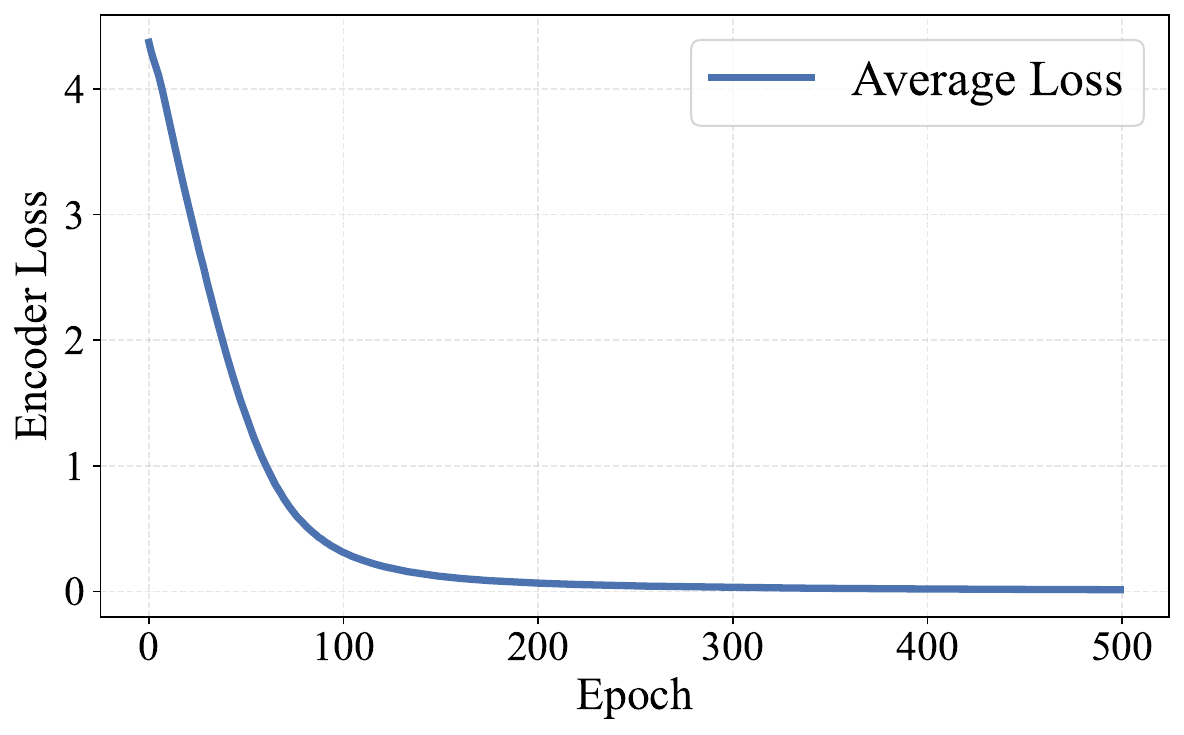}
\caption{Loss curve of our encoder}
\end{subfigure}\hfill
\begin{subfigure}{0.50\linewidth}
\centering
\includegraphics[width=\linewidth]{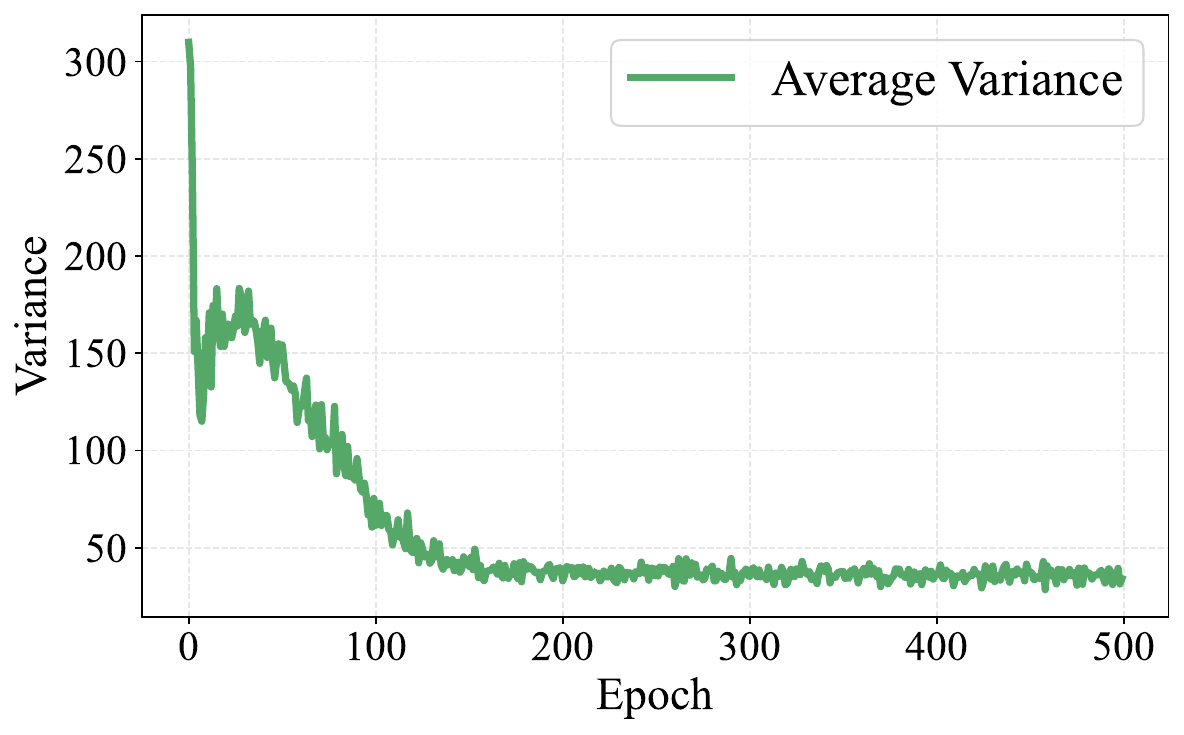}
\caption{Average variance curve}
\label{Fig:humanoid_score}
\end{subfigure}
\caption{(a) The loss curve of the classification-based encoder. (b) The average variance curve of the number of randomly sampled embeddings from the latent space located in the vicinity of each $u$.}
\label{Fig:curve}
\end{figure}

We examine the effectiveness of leveraging the classification-based encoder in our framework to achieve a uniform distribution of feature means $\mathcal{U}=\{u_1, u_2, ..., u_n|u_i=\mathcal{C}(m_i)\}$, where $\mathcal{U} \subset \mathcal{Z}$ on the surface of a high-dimensional hypersphere. After training the classifier $C$, all $u \in \mathcal{U}$ are considered to have been distinctly separated according to Neural Collapse. L2-normalized samples from a multivariate Gaussian distribution approximate uniform sampling on the unit sphere in high-dimensional spaces due to the concentration of measure and the isotropy of the Gaussian distribution \cite{vershynin2018high, blum2005random}. We conducted an approximating uniformly random sampling of $1000 \times n$ points from the unit hypersphere, i.e., $z=\overline{z}/||\overline{z}||$, $ \overline{z} \sim \mathcal{N}(0, I)$. The proximity of each sampled latent to the $u$ is determined using cosine distance. We then quantified the number of embeddings in the vicinity of each $u$ and calculated the variance of these counts. A lower average variance suggests a more uniform distribution of the set \( \mathcal{U} \) across the hypersphere. As shown in Figure~\ref{Fig:nc}, for \textit{MSA} dataset, the distribution of points around each embedding in our approach demonstrates remarkable consistency, with a low variance of 31 for all $u$. Figure~\ref{Fig:curve} shows how the average variance changes during the training of our encoder on \textit{Sword\&Shield} dataset. As the encoder loss decreases, the motion representation becomes more uniformly distributed across the spherical space. In contrast, a distribution derived from applying a metric learning loss to maximize the mutual distances between embeddings exhibits significant disparities in the density of points surrounding each embedding, resulting in a considerably higher variance of 500. These results prove the efficacy of our design.

Another observed phenomenon for our encoder is that although skill embeddings are learned through a classification problem and are uniformly mapped onto a spherical latent space, some semantically connected motions exhibit relatively smaller cosine distances to each other compared to other motions that are more distinct. 
This phenomenon arises despite the uniform distribution in the latent space, due to the geometric properties of the sphere, where points that are semantically closer tend to have smaller angular separations, even though the overall arrangement remains evenly distributed. 
For example, in \textit{Sword\&Shield} dataset, the five motions closest to the feature mean of the WalkRight01 motions are Standoff\_Circle, Counter\_Atk02, RunRight, and ShieldKnock. Among these, Standoff\_Circle and RunRight share a similar semantic meaning with WalkRight01, as they all involve the character moving to the right. Similarly, the five motions nearest to the feature mean of the 4xCombo01 are 3xCombo05, 4xCombo03, 3xCombo04, WalkBackward02, and ParryBackward01. The first three, 3xCombo05, 4xCombo03, and 3xCombo04, all involve consecutive sword attacks, while WalkBackward02 and ParryBackward01 represent motions where the character moves backward. This phenomenon becomes more apparent when applied to more extensive and diverse datasets. For the Move\_Run action, for instance, the five closest motions are Running\_Backward, Trot, Fast\_Walking, Jogging, and Sneak\_Moving. These actions are all locomotion skills, showing closer proximity in the latent space compared to more distinct actions like dancing or sports-related movements. However, for other categories consisting of more complex motions, this characteristic is less evident. We hypothesize that the complexity and diversity of human movements make it challenging for the model to organize them intuitively based on motion similarity, where motions may share semantic similarities but differ significantly in their actual physical movement.

\subsection{Ablation Study}
Our approach integrates two key techniques: a classification-based encoder that ensures the motion clips from the dataset are uniformly distributed in the latent space, and the Embedding Expansion technique that generates skill embedding clusters for each motion clip, providing expressive variations in the stylized embeddings. To test the effectiveness of our design, we conducted two ablation studies. In experiment 1, we only used the classification-based encoder to distribute the features of each motion clip uniformly across the latent space, without applying Embedding Expansion. Conditional imitation learning was then performed using these mapped features. In experiment 2, we do not use the encoder for uniform distribution and instead randomly sampled points on a spherical surface to serve as the feature means for each motion in the dataset. Then the Embedding Expansion is applied to these sampled features. After training, we sampled 512 embeddings from the trained latent space and applied them iteratively to the controller for 600 steps. The generated actions were evaluated using Equation~\ref{eq:recon}. In Figure~\ref{Fig:reconstruc}, we compare the performance of these two models on the \textit{Sword\&Shield} and \textit{MSA} datasets. On the \textit{Sword\&Shield} dataset, the non-uniform model achieved a reconstruction score above 0.5 for 72.7\% of frames across all motion clips, and the model without Embedding Expansion achieved 60.8\%. In contrast, our model combining uniformly distributed class centers with Embedding Expansion achieved 91.2\%. A similar trend was observed on the \textit{MSA} dataset, where the three models achieved 75.7\%, 54.1\%, and 94.3\%, respectively. These results demonstrate the effectiveness of our approach, showing that the uniformly distributed class centers and the Embedding Expansion technique collectively enable our versatile, adaptable, and controllable policy.

\section{Discussion}
In this paper, we introduce a novel framework that enables physically simulated characters to learn uniformly distributed embedding clusters for each skill. Our approach leverages a hypersphere as the embedding space to create uniformly distributed clusters corresponding to each skill. We begin by learning a classification-based encoder to map the distinctive features of each reference motion onto a high-dimensional sphere, utilizing the Neural Collapse phenomenon to achieve uniform distribution. Next, we employ conditional imitation learning, combined with the proposed Embedding Expansion method, to construct a von Mises-Fisher distribution where each unique feature serves as the mean direction, and the concentration parameter controls the distribution's sharpness. After training, embeddings within each cluster generate behaviors that align with the corresponding skill, offering high diversity while maintaining consistency with the behavioral style represented by the motion clip. Furthermore, for each skill cluster center, synchronizing motion-progress-embedded states with the timing of the respective motion at each step allows the policy to execute the full specified skill, producing movements identical to the reference motion, which is particularly valuable for interactive control. However, compared to other skill-conditioned models, our approach, where each skill cluster is modeled using a von Mises-Fisher (vMF) distribution, has certain limitations. Specifically, when drawing random samples from the continuous skill embedding space for a given skill, it is possible for the sample to be far from the cluster center and closer to another skill cluster or even lie at the intersection of two clusters. This may result in motions that differ from those intended for the specified skill. Such behavior could impact users requiring precise motion generation for specific actions. We aim to address this issue in future work. Additionally, we will further investigate the effects of the well-structured latent space constructed by this method on high-level policy.

\bibliographystyle{eg-alpha-doi} 
\bibliography{egbibsample}

\end{document}